\documentclass[12pt]{article}
\usepackage[dvips]{graphicx}
\usepackage{pdproc}

  \makeatletter 
  \def\@cite#1{[#1]} 
  \makeatother    
\begin{document}

\renewcommand{\thefootnote}{\alph{footnote}}

\title{
Suppressed neutrino oscillations and 
large lepton asymmetries
}

\author{A.D. DOLGOV$^{(a)(b)(c)}$ and 
FUMINOBU TAKAHASHI$^{(d)}$
}

\address{ 
$^{(a)}${\it INFN, sezione di Ferrara,
Via Paradiso, 12 - 44100 Ferrara,
Italy} \\
$^{(b)}${\it ITEP, Bol. Cheremushkinskaya 25, Moscow 113259, Russia.
}  \\
$^{(c)}${\it ICTP, Trieste, 34014, Italy
}  \\
$^{(d)}$Institute for Cosmic Ray Research, \\
University of Tokyo, Kashiwa 277-8582, Japan
\\ {\rm E-mail: fumi@icrr.u-tokyo.ac.jp}}

\abstract{
It is shown that hypothetical neutrino-majoron coupling can suppress
neutrino flavor oscillations in the early universe, in contrast to 
the usual weak interaction case. This reopens a window for a noticeable 
cosmological lepton asymmetry which is forbidden for the large mixing 
angle solution in the case of standard interactions of neutrinos. }

\normalsize\baselineskip=15pt

\section{Introduction \label{s-intr}}

Cosmological lepton asymmetry is not directly measurable, in contrast
to baryon asymmetry, but may be observed or restricted through its
impact on big bang nucleosynthesis (BBN), large scale structure
formation, and the angular spectrum of the cosmic microwave background
radiation (CMBR), for a review see e.g. Ref.~\cite{dolgov02}.  At the
present time the best bounds follow from the consideration of
BBN.  According to Ref.~\cite{hansen01} they are:
$|\xi_e|<0.2$ and $|\xi_{\mu,\tau}|<2.6$. Therefore the lepton asymmetry
can be large, and its origin and implications are discussed by many 
authors~\cite{largeL_smallB}.

The bounds on chemical potentials of $\nu_\mu$ and $\nu_\tau$ can be
significantly improved because of the strong mixing between different
neutrino flavors~\cite{nu-mixing}. This mixing gives rise to the fast
transformation between $\nu_e$, $\nu_\mu$, and $\nu_\tau$ in the early
universe and leads to equilibration of asymmetries of all neutrino
species. Thus the BBN bound on any chemical potential becomes
essentially that obtained for $\nu_e$~\cite{act-act} (see also the
papers~\cite{lunardini01}):
\begin{equation}
|\xi_{e,\mu,\tau}| < 0.07.
\label{xiemt}
\end{equation}
In this case the cosmological impact of neutrino degeneracy would be
negligible. 

It is interesting to see if one could reasonably modify the standard
model to allow large muonic and/or tauonic charge asymmetries,
together with a small electronic asymmetry, to avoid conflict with
BBN. This is the aim of this work. A natural generalization is to
introduce an additional interaction of neutrinos with massless or
light (pseudo)Nambu-Goldstone boson, majoron~\cite{majoron}. 
 Let us note
that in this paper we consider an impact of neutrino majoron
interactions on the oscillations between active neutrinos and not on
active-sterile oscillations~\cite{babu92,bento01}.

\section{Neutrino-majoron interactions \label{s-eff-pot}}
We assume the following neutrino-majoron interaction:
\begin{eqnarray}
{\cal L} &=&- \frac{1}{2} \partial_\mu \chi \partial^\mu \chi - \sum_a
\overline{\nu}_a \gamma^\mu \partial_\mu \nu_a +
 \frac{i}{2} \chi \left( g_{ab}\, \nu_a ^T C \nu_b +g_{ab}^* \,
 \nu_b^\dagger C \nu_a^* \right),
 \label{eq:maj-nu-int}	
\end{eqnarray}
where $\chi$ is the majoron field, and $\nu_a$ is four-component
representation of neutrino of flavor $a$. Here 
 $\nu_a$ is taken to be left-handed. 
This interaction induces the effective potential for neutrinos with
momentum ${\bf p}$~\cite{Dolgov:2004jw}:
\begin{equation}
\left[V_{\bf p}^{(\chi)}\right]_{ab} = \int \frac{d^3 {\bf q}}{(2 \pi)^3}
~~\frac{1}{4 \left|{\bf p} \right| \left|{\bf q} \right|} \left[
g^\dagger \left(\rho^T_{\bf q} + \overline{\rho}^T_{\bf q} 
+f_{\chi}({\bf q})\cdot {\bf 1}
\right) g
\right]_{ab},
\label{eq:eff_potential_total}
\end{equation}
where 
$\rho_{\bf p}~(\bar \rho_{\bf p})$ is the density matrix for
(anti)neutrinos, $f_{\chi}({\bf p})$ is the 
 number density of 
majorons with momentum $\bf{p}$,  and ${\bf 1}$ is the unit matrix in the flavor basis.

The weak interaction as well induces the effective potential for neutrinos,  $V^{(w)}$,
the exact form of which can be found in e.g. Ref.~\cite{Sigl:fn}. If $V^{(\chi)}$ dominates
over $V^{(w)}$, the neutrino oscillations can be suppressed. To be precise,
the diagonal part of the potential
$V^{(\chi)}_{aa}$ should be larger than the weak potential $V^{(w)}$,
 while its off-diagonal components must be
much smaller than the diagonal ones, that is, the flavor symmetry in
the neutrino-majoron interactions should be strongly broken. To this end,
we assume that the coupling constant matrix $g_{ab}$ is approximately diagonal 
and one of the diagonal components dominates over the other components. 
For a more generic form of $g_{ab}$, see Ref.~\cite{Dolgov:2004jw}.

The coupling constants $g_{aa}$ should not be too large,
otherwise flavor non-conserving reactions of the type 
$\nu_e\,\nu_a \leftrightarrow \bar \nu_e \bar\nu_a$ (or similar) would lead
to equilibration of all leptonic charges. To avoid that the rate 
of these reactions, $\Gamma_{ea} \sim \sigma_{ea} T^3$, should be
smaller than the cosmological expansion rate $H\sim T^2/m_{Pl}$, 
where $m_{Pl} = 1.221\cdot 10^{22}$ MeV is the Planck mass. Thus,
to suppress $e-\mu$ or $e-\tau$ transformation through direct 
reactions one needs
\begin{equation}
g_{aa}^2g_{ee}^2 <10^{-22} \left(\frac{T}{1{\rm~ MeV}}\right).
\label{g-react}
\end{equation}
This conditions should be satisfied for temperatures above the BBN
range, i.e. $T>1$ MeV. Similarly, if we require that $\nu_a\,\nu_a
 \leftrightarrow \bar \nu_a \bar\nu_a$ should not occur efficiently, the coupling
constants must satisfy a similar inequality with $g_{ee}$ replaced
with $g_{aa}$.

Furthermore, there are quite strong limits on possible coupling of majoron to
neutrinos which follow from astrophysics. Astrophysics allows either very small or quite
large coupling constants. The former is quite evident, while the
latter appears because strongly interacting majorons, though
efficiently produced inside a star, cannot propagate out and carry
away the energy, thus opening a window for large values of the
coupling. It is not so for the coupling to $\nu_e$ because the latter
is bounded from above by the data on double beta decay,
$g_{ee}< 3\cdot 10^{-5}$.  Together with the supernova bounds, the
upper limit is shifted down to $g_{ee} < 4\cdot
10^{-7}$~\cite{farzan02}, with a small window around $(2-3)\cdot
10^{-5}$. So we assume in the following that $g_{ee}  \ll 
10^{-7}$. For $\mu$ or $\tau$ the allowed regions are: $g_{aa} <
(3-5)\cdot 10^{-6}$ or $g_{aa}>(3- 5)\cdot 10^{-5}$. Not to erase the
lepton asymmetries, the former allowed region is assumed.

\section{Results}
\newcommand{\gtrsim}{ \mathop{}_{\textstyle \sim}^{\textstyle >} }
Using the effective potential shown in the previous section,
we have calculated the evolution of the lepton asymmetries
both analytically and numerically. Here we show only
the numerical results, and see Ref.~\cite{Dolgov:2004jw}
for the analytical method. 
In doing the numerical calculations, we have assumed
that the mixing is effective only between two neutrinos
since the atmospheric neutrino mass difference is much
larger than the solar one. Also the coupling constant matrix
$g_{ab}$ is  approximated to be $g_{ab} = g \,\delta_{a \mu'}
\delta_{b \mu'}$.  
The numerical result is shown in Fig.~1, which says that, 
 as $|g|$ increases, the oscillations become less efficient
and completely stop for $|g| \gtrsim 10^{-7}$. Note that
we have obtained consistent results by the analytic method.

 \section{Conclusions}
In this paper we have shown that the hypothetical neutrino-majoron
interaction can suppress neutrino oscillations in the primordial
plasma to prevent lepton asymmetries of all neutrino species from
being equilibrated. The exact form of the effective potential induced
by this interaction is calculated. We have found an allowed range of
the coupling constant: $10^{-7} < |g| < 5\cdot 10^{-6}$, which satisfies the astrophysical
bounds and makes the scenario operative.  For the coupling constant in
this range, $\nu_e-\nu_{\mu'}$ oscillation in the early Universe is
blocked, thereby keeping the cosmological lepton asymmetry of electron
type unchanged.  The upper bound comes from the requirement that
lepton number is effectively conserved, and the lower bound is
obtained from the study of the evolution of the lepton asymmetries
both analytically and numerically, in two flavor approximation.  The
constant matrix in the simplest class of majoron models can satisfy
the desired constraints, in the case of the normal mass hierarchy.
Thus we conclude that an addition of the
majoron field to the standard model can reopen a possibility that the
effect of $\xi_e$ is compensated by large $\xi_{\mu,\tau}$ (or by the
extra energy of majoron itself), thereby curing a probable discrepancy
between the BBN and CMBR.

\begin{figure}[htb]
\begin{center}
\includegraphics*[width=10cm]{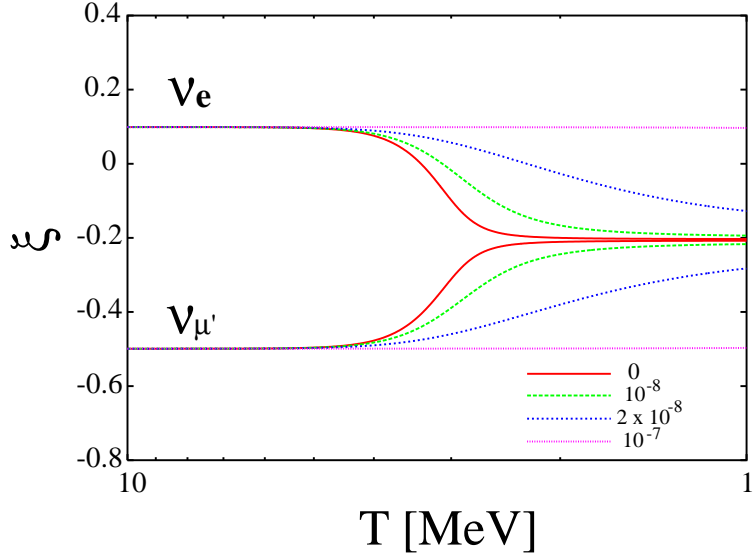}
\caption{%
The evolutions of $\xi_e$ and $\xi_{\mu'}$ for several values of $|g|$
with $\sin^2 \theta = 0.315$ and $\delta m_{21}^2 = 7.3 \times 10^{-5} {\rm eV}^2$. 
The initial conditions are $\xi_e = 0.1$ and $\xi_{\mu'}=-0.5$.
}
\label{fig1}
\end{center}
\end{figure}

\bibliographystyle{plain}

\end{document}